\documentclass[prd,aps,twocolumn,preprintnumbers,nofootinbib,showpacs,amsmath,amssymb,superscriptaddress]{revtex4-1}
\usepackage{dcolumn}
\usepackage[margin=2.4cm]{geometry}
\usepackage{graphicx}
\usepackage{epsfig}
\usepackage{amsmath}
\usepackage{amssymb}
\usepackage{float}
\usepackage{cancel}
\usepackage{amsfonts}
\usepackage{enumitem}
\usepackage[font={footnotesize,it}]{caption}
\usepackage[titletoc,toc,title]{appendix}
\usepackage{hyperref}
\usepackage{epsfig}
\usepackage{cancel}
\usepackage{enumitem}
\usepackage{subcaption}
\usepackage{lipsum}
\usepackage[utf8]{inputenc}
\hypersetup{
    colorlinks=true,
    linkcolor=blue,
    filecolor=red,
    urlcolor=blue,
    citecolor=red
} 

\newcommand{\Tr}{\text{tr~}}

\begin{document} 

\title{\boldmath An all order exact result for the anomalous dimension of the scalar primary in Chern Simons Vector Models}

\author{Sachin Jain}
\email{ sachin.jain@iiserpune.ac.in }
\affiliation{Indian Institute of Science Education and Research, Homi Bhabha Rd, Pashan, Pune 411 008, India}\author{Vinay Malvimat}
\email{ vinaymm@iiserpune.ac.in }
\affiliation{Indian Institute of Science Education and Research, Homi Bhabha Rd, Pashan, Pune 411 008, India}
\author{Abhishek Mehta}
\email{ abhishek.mehta@students.iiserpune.ac.in }
\affiliation{Indian Institute of Science Education and Research, Homi Bhabha Rd, Pashan, Pune 411 008, India}
\author{Shiroman Prakash}
\email{ sprakash@dei.ac.in }
\affiliation{Dept. of Physics and Computer Science, Dayalbagh Educational Institute, Agra, India} 
\author{Nidhi Sudhir}
\email{ nidhi.sudhir@students.iiserpune.ac.in }
\affiliation{Indian Institute of Science Education and Research, Homi Bhabha Rd, Pashan, Pune 411 008, India}

% The "\note" macro will give a warning: "Ignoring empty anchor..."
% you can safely ignore it.

\begin{abstract}
We present a conjecture for the leading $1/N$ anomalous dimension of the scalar primary operator in $U(N)_k$ Chern-Simons theories coupled to a single fundamental field,
to all orders in the t'Hooft coupling $\lambda=\frac{N}{k}$. Following this we  compute the anomalous dimension of the scalar in a Regular Bosonic theory perturbatively at  two-loop order and demonstrate that matches exactly with the result predicted by our conjecture. We also
show that our proposed expression for the anomalous dimension is consistent with all other existing two-loop perturbative results, which constrain its form at both weak and strong 
coupling thanks to the bosonization duality. Furthermore, our conjecture passes a novel non-trivial all loop test which provides a strong evidence for its consistency. 
%In case our conjecture is not correct, it can be thought of 
%as the leading term in a two-sided Pad\'e approximation that takes into account 
%data perturbative data at weak and strong coupling.
\end{abstract}

\maketitle
\flushbottom

\section{Introduction}

$U(N)_k$ %and $O(N)_k$ 
Chern-Simons theories coupled to a single fundamental field are an important class of conformal field theories that are solvable in the large $N$ limit \cite{Giombi:2011kc, Aharony:2011jz}. As emphasized in \cite{MZ, Aharony:2012nh,Skvortsov:2018uru} four such theories exist depending on the choice of fundamental matter, which can be divided into two classes: quasi-fermionic and quasi-bosonic. The quasi-fermionic class includes the theory with one species of fundamental fermions as matter and the theory with critical (Wilson-Fisher) bosons as matter. Both these theories are believed to be related by a strong-weak coupling duality \cite{MZ, Aharony:2012nh,Skvortsov:2018uru, GurAri:2012is, Aharony:2012ns, Gur-Ari:2015pca, Bedhotiya:2015uga, Minwalla:2015sca, Aharony:2015mjs}. The quasi-bosonic class includes the theory with matter as (non-critical) bosons and the theory with critical (Gross-Neveu) fermions as matter. Again, both these theories are related by strong-weak duality, discussed extensively in \cite{Aharony:2018pjn}. See also, e.g., \cite{Yokoyama:2016sbx, Seiberg:2016gmd, Murugan:2016zal, Kachru:2016rui, Radicevic:2016wqn, Hsin:2016blu, Jain:2013py,  Jain:2013gza, Jain:2014nza} for additional tests and discussion of the bosonization duality.  

An important feature of these theories is that they contain a very sparse spectrum of single-trace primary operators. There is exactly one single-trace primary operator for each spin $s$, which we denote as $j_s$. When 't Hooft coupling $\lambda=0$, these currents are exactly conserved, and therefore have scaling dimensions given by the unitarity bound $\Delta_s= s+1$ for nonzero $s$. As argued in \cite{Giombi:2011kc, Aharony:2011jz}, a simple argument based on conformal representation theory implies that the scaling dimensions of these currents are protected in the large $N$ limit, even when $\lambda \neq 0$. The leading corrections to the scaling dimensions are proportional to $1/N$: $\Delta_s = (s+1) + \gamma_s(\lambda)+O(\frac{1}{N^2})$ where $\gamma_s(\lambda)$ corresponds to the anomalous dimension of spins-$s$ primary operator at order $1/N$. For operators with spin $s\neq 0$, the scaling dimensions can be determined from planar three-point functions using the slightly broken higher-spin symmetry \cite{MZ} of the theory \cite{Giombi:2016zwa}. 

The results and methods of \cite{Giombi:2016zwa} rely on slightly-broken higher-spin symmetry \cite{Giombi:2016hkj, Nii:2016lpa, Skvortsov:2015pea}, and are valid only for $s\neq 0$. Although it is possible to analytically continue the formulas derived in \cite{Giombi:2016zwa}, this gives us a result that is inconsistent with perturbative computations (which are possible at both weak and strong coupling thanks to the bosonization duality). Hence the leading $1/N$ correction to the anomalous dimension of the scalar primary $j_0$ remains unknown at present. This quantity is interesting for a variety of reasons. As argued in \cite{Aharony:2018pjn}, it plays an important role in determining the fixed point for the $\phi^6$ coupling in the quasi-bosonic family of theories at $1/N$. 

Quite interestingly, in condensed matter physics, the scaling dimension of $j_0$ is extremely significant as it determines an experimentally-measurable critical exponent for certain quantum Hall phase transitions \cite{WenWu, Mott,Hui:2017pwe, Hui:2017cyz}. However, for comparison with experiments one would require  finite and small $N$. In this context, the anomalous dimension of the scalar primary is one of the simplest physical observables for our theory, and it is rather striking that it remains unknown.

In principle, an exact Feynman diagram calculation could be performed to calculate the anomalous dimension of $j_0$ to all orders in $\lambda$ using the light-cone gauge, (as discussed in \cite{Gurucharan:2014cva}) but this is not a possibility at present for what appear to be insurmountable technical reasons. In particular, one of the crucial ingredients, the exact ladder diagram \cite{ Aharony:2012nh, Bedhotiya:2015uga}, is not known off-shell.

Here, motivated by the results \cite{Giombi:2016zwa} for $\gamma_s$, and perturbative calculations, including a new a calculation of the two-loop anomalous dimension of the scalar primary in the theory coupled to fundamental bosons, we  conjecture a simple all-orders expression for the anomalous dimension $s=0$. We will show that this conjecture passes several non-trivial consistency checks.

The article is organized as follows. In section \ref{S2} we briefly discuss the parameters in the quasi-bosonic and quasi fermionic theory and setup our notations for those parameters. In section \ref{QBS1} we perform a perturbative computation of the anomalous dimension for the scalar in the regular bosonic theory and also describe the result for the critical fermionic theory in the literature. Following this, in section \ref{QFT} we describe the results for the anomalous dimension in quasi-fermionic theories. Following this, in section \ref{ALT} we briefly review a result obtained in   \cite{Aharony:2018pjn} which serves as an all loop test for our proposal. In section \ref{kr} we demonstrate that the naive analytic continuation of the spins-$s$ operator to $s=0$ fails to reproduced the correct anomalous dimension for all the theories. Subsequently in section \ref{Conj} we propose our conjecture for the anomalous dimension of the scalar operators in both quasi-fermionic and quasi-bosonic and demonstrate that our conjecture reproduces all known perturbative results and also passes a non-trivial all loop test.
In Appendix we will argue that our conjecture can also  be thought of as a two-sided Pad\'e approximation, which makes use of perturbative data at both weak and strong coupling, in the spirit of S-dualty improved perturbation theory \cite{Sen:2013oza}. Of course, this is only possible because of the bosonization duality. 

%As mentioned earlier, the scaling dimension of $j_0$ determines a critical exponent in quantum Hall phase transitions. While comparison to the experiment requires a finite $N$ calculation, which is not possible at present, some condensed matter authors have tried to obtain estimates for this critical exponent using with a large number of flavours \cite{WenWu, Mott,Hui:2017pwe}. The large flavour limit used in condensed matter is a qualitatively different kind of large $N$ limit compared to the large colour limit that we are investigating in this paper, and it is interesting to compare our estimates for the critical exponent obtained from our expression for $\Delta_0$ to those in the literature.

\section{Parameters and Theories}\label{S2}
Let us carefully review the theories under study and their relations via RG flow and bosonization duality. 

The quasi-bosonic family of theories flows to the quasi-fermionic family of theories under RG flow. In \cite{MZ}, the quasi-bosonic family is described by three parameters\footnote{The analysis of \cite{MZ} is valid for theories with only even spins, e.g. $O(N)$ vector models. For the $U(N)$ vector models which we study here, the analysis of \cite{MZ} has not been carried out, and there may be additional parameter, corresponding to the strength of an additional Chern-Simons $U(1)$ Chern-Simons field that could couple to the spin-1 conserved current, which we assume is turned off here.} $\tilde{\lambda}_{QB}$, $\tilde{N}_{QB}$ and $\tilde{\lambda}_{6,QB}$; and the quasi-fermionic family is described by two parameters $\tilde{\lambda}_{QF}$ and $\tilde{N}_{QF}$. The parameter $\tilde{N}$ is defined via the two-point function of the stress-energy tensor, and is a measure of the number of degrees of freedom of each theory -- we will only be interested in the large $\tilde{N}$ limit and the first non-trivial $1/\tilde{N}$ corrections. In this limit, the spectrum is independent of the parameter $\tilde{\lambda}_{6,QB}$ so we will ignore it in the discussion that follows.

The celebrated bosonization duality states that each family of theories has two very-different-looking descriptions. The quasi-bosonic family can be described as a theory of $N_b$ complex bosons transforming in the fundamental representation of $U(N_b)$, coupled to a level-$\kappa_b$ Chern-Simons gauge field. It can also be described as a theory of $N_f$ Dirac ``critical'' fermions, in the fundamental representation of $U(N_f)$ coupled to a level $\kappa_f$ Chern-Simons gauge field. The quasi-fermionic family can be described as a theory of $N_b$ critical complex bosons transforming in the fundamental representation of $U(N_b)$, coupled to a level-$\kappa_b$ Chern-Simons gauge field. It can also be described as a theory of $N_f$ Dirac fermions, in the fundamental representation of $U(N_f)$ coupled to a level $\kappa_f$ Chern-Simons gauge field. 

This duality is well-tested in the large $N_{b/f}$ limit, with $\lambda_{b/f} \equiv \frac{N_{b/f}}{k_{b/f}}$ held fixed. In this limit we have the following relation between the parameters:
\begin{eqnarray}
\tilde{N}_{QB} & = & 2 N_b \frac{\sin (\pi \lambda_b)}{\pi \lambda_b} = 2 N_f \frac{\sin (\pi \lambda_f)}{\pi \lambda_f}\nonumber \\
\tilde{N}_{QF} & = & 2 N_b \frac{\sin (\pi \lambda_b)}{\pi \lambda_b} = 2 N_f \frac{\sin (\pi \lambda_f)}{\pi \lambda_f} \label{par1}\\
\tilde{\lambda}_{QB} & = &  \tan \left(\frac{\pi \lambda_b}{2} \right) = \cot \left(\frac{\pi \lambda_f}{2} \right)\nonumber  \\
\tilde{\lambda}_{QF} & = &  \cot \left(\frac{\pi \lambda_b}{2} \right) = \tan \left(\frac{\pi \lambda_f}{2} \right)\nonumber 
\end{eqnarray}
Because $N_{b/f}$ and $\kappa_{b/f}$ are integers (or half-integers), the parameters $\lambda_{b/f}$ and $N_{b/f}$ do not run under RG flow from quasi-bosonic theory to quasi-fermionic theory. Under RG flow, the quasi-bosonic theory defined by $\tilde{\lambda}_{QB}$ and $\tilde{N}_{QB}$ flows to the quasi-fermionic theory described by:
\begin{eqnarray}
\tilde{\lambda}_{QF} & = & \frac{1}{\tilde{\lambda}_{QB}} \\
\tilde{N}_{QF} & = & \tilde{N}_{QB}.
\end{eqnarray}
We henceforth use $\tilde{N}$ without any subscript.

Let us denote the scaling dimension of the scalar primary $j_0$ in the quasi-bosonic theory as $\Delta_0$, and the scaling dimension of $\tilde{j}_0$ in the QF theory as $\tilde{\Delta}_0$. We define the anomalous dimension as:
\begin{equation}
\Delta_0 = 1 + \gamma_0,~~~ \tilde{\Delta}_0= 2 + \tilde{\gamma}_0
\end{equation}
\section{Quasi-Bosonic Theories}\label{QBS1}
\subsection{Perturbative Computations in the Regular Bosonic Theory}
In this section, we begin by computing the anomalous dimension of $j_0=\bar{\phi}\phi$ in the regular bosonic theory, i.e., $SU(N)_k$ Chern Simons theory coupled to a single complex scalar field, to two loops. (The leading $1/N$ correction to the anomalous dimension is the same whether one considers the $U(N)$ or $SU(N)$ theories, although subleading  corrections may differ.) This will serve as a non-trivial check for our conjecture. Our computation closely follows the calculation of the anomalous dimension of $j_0$ in the $O(N)$ theory carried out in \cite{Aharony:2011jz}. All our calculations in this appendix are in the bosonic theory, so we drop the subscript $b$ in what follows.

We also refer to related perturbative computations in Chern-Simons theory which appear in \cite{Avdeev:1991za, Ivanov:1991fn, Chen:1992ee, Banerjee:2013nca}.
To this end, we calculate the anomalous dimension of the operator $j_0$ at two loops. The diagrams which we need to evaluate are given in figure \ref{B1f},\ref{UVFin},\ref{gh}. Our conventions, Feynman rules and gauge choices are provided in the appendix \ref{fey} \\

\begin{figure}[h]
    \includegraphics[scale=0.4]{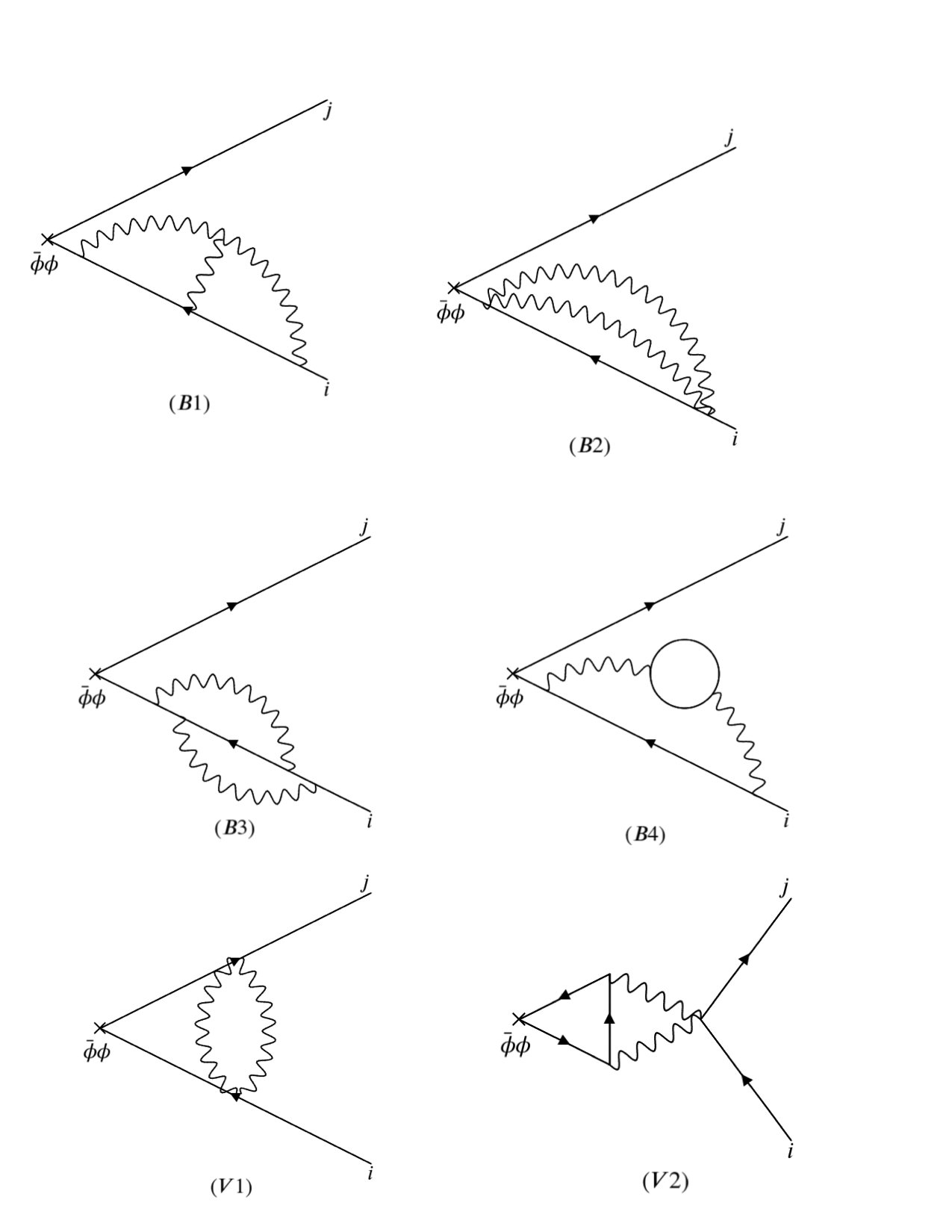}
    \caption{Diagrams ($B1$)-($B4$) depict loop corrections to the propagator. Diagrams (V1) and (V2) are loop corrections to the vertex.}
    \label{B1f}
    \centering
\end{figure}The  logarithmic divergences arising due to the loop correction of the propagators depicted in the diagrams ($B1$)-($B4$) of the figure \ref{B1f} are given by
\begin{eqnarray}
(B1) &=& \frac{2}{3 k^2}C_2 C_3 ~\log[\Lambda]=\frac{1-N^2}{3k^2}~\log[\Lambda]\label{B11}\\
(B2) &=&\frac{2}{3 k^2}(C_3^2+\frac{C_2C_3}{4})~\log[\Lambda]=\frac{ N^4 - 3 N^2 +2}{12 k^2N^2}\log[\Lambda]\notag\\
(B3) &=&\frac{8}{3 k^2}(C_3^2+\frac{C_2C_3}{2})\log[\Lambda]=\frac{2}{3 k^2} (\frac{1}{N^2}-1)\log[\Lambda]\notag\\
(B4) &=& \frac{4}{3 k^2} C_1 C_3 \log[\Lambda]=\frac{1}{3 k^2} \left(N-\frac{1}{N}\right)\log[\Lambda]\notag
\end{eqnarray}
The logarithmic divergences arising from the corrections to the vertex depicted in figure \ref{B1f} are given by
\begin{eqnarray}
(V1) &=&\frac{4}{ k^2}(C_3^2+\frac{C_2C_3}{4})~\log[\Lambda]=\frac{ N^4 - 3 N^2 +2}{2~ k^2N^2}\log[\Lambda]\notag\\
(V2) &=& \frac{8}{ k^2}C_1 C_3 ~\log[\Lambda]=\frac{2}{k^2} \left(N-\frac{1}{N}\right)\log[\Lambda]\label{V22}.
\end{eqnarray}
Following  \cite{Gurucharan:2014cva}, we use   these results, to compute the $\mathcal{O}(\frac{1}{N})$ logarithmic divergence of the two-point function $\langle j_0 j_0 \rangle$ to be:
\begin{equation}
2(B1+B2+B3+B4)+V1+V2=\frac{8}{3}\frac{\lambda^2}{N}\log[\Lambda] +O(\frac{1}{N^2}),
\end{equation}
where we have used re-expressed the result in terms of $\lambda\equiv \frac{N}{k}$. Note that the $U(N)$ result is just the large-$N$ limit of the $SU(N)$ result.\\

Note that the loop corrections to the propagator should be taken on each of the two legs of the vertex diagrams ($B1$)-($B4$) depicted in figure \ref{B1f} and hence they contribute twice to the two-point function. 

Now we briefly describe how to obtain the anomalous dimension of operator $j_0$ from two point function of the same operator.The two-point correlation function of the scalars in a $d$-dimensional CFT in momentum space is given by
\begin{equation}
\label{2ptfn}
\langle j_0(p) j_0(0) \rangle=\frac{c_1}{p^{2\Delta-d}}.
\end{equation}
The scaling dimension $\Delta$ can be expressed in $\frac{1}{N}$ expansion as 
\begin{equation}
\label{Nanex}
\Delta=\Delta_0+\gamma_0+O(\frac{1}{N^2})
\end{equation}
where  $\Delta_0$ is
classical scaling dimension, $\gamma_0$ is anomalous dimension to order $\frac{1}{N}$.
Plugging \eqref{Nanex} in \eqref{2ptfn} and expanding to leading order around  $\gamma_0=0$, we obtain
\begin{eqnarray}
\langle j_0(p) j_0(0) \rangle &=& \frac{1}{p^{2\Delta_0-d}}(1-2 \gamma_0 \log{p})]
\end{eqnarray}
Hence the anomalous dimension is given by $-1/2$ times the logarithmic divergence we obtained earlier.  Keeping  corrections in the anomalous dimension upto $\mathcal{O}[\frac{1}{N}]$, this leads us to the following expression for the anomalous dimension at  $\mathcal{O}[\lambda_b^2]$ 
\begin{equation}\label{anmcal}
\gamma_0 = \frac{1}{N_b} \left( -\frac{4}{3} \lambda_b^2+ O(\lambda_b^4)\right).
\end{equation}
In a subsequent section, we will demonstrate that the  anomalous dimension in Eq.\eqref{anmcal} matches exactly with the  perturbative expansion of our conjectured expression in \eqref{gt0}.

\subsection{UV Finite diagrams}
\begin{figure}[h]
\centering
 \includegraphics[scale=0.34]{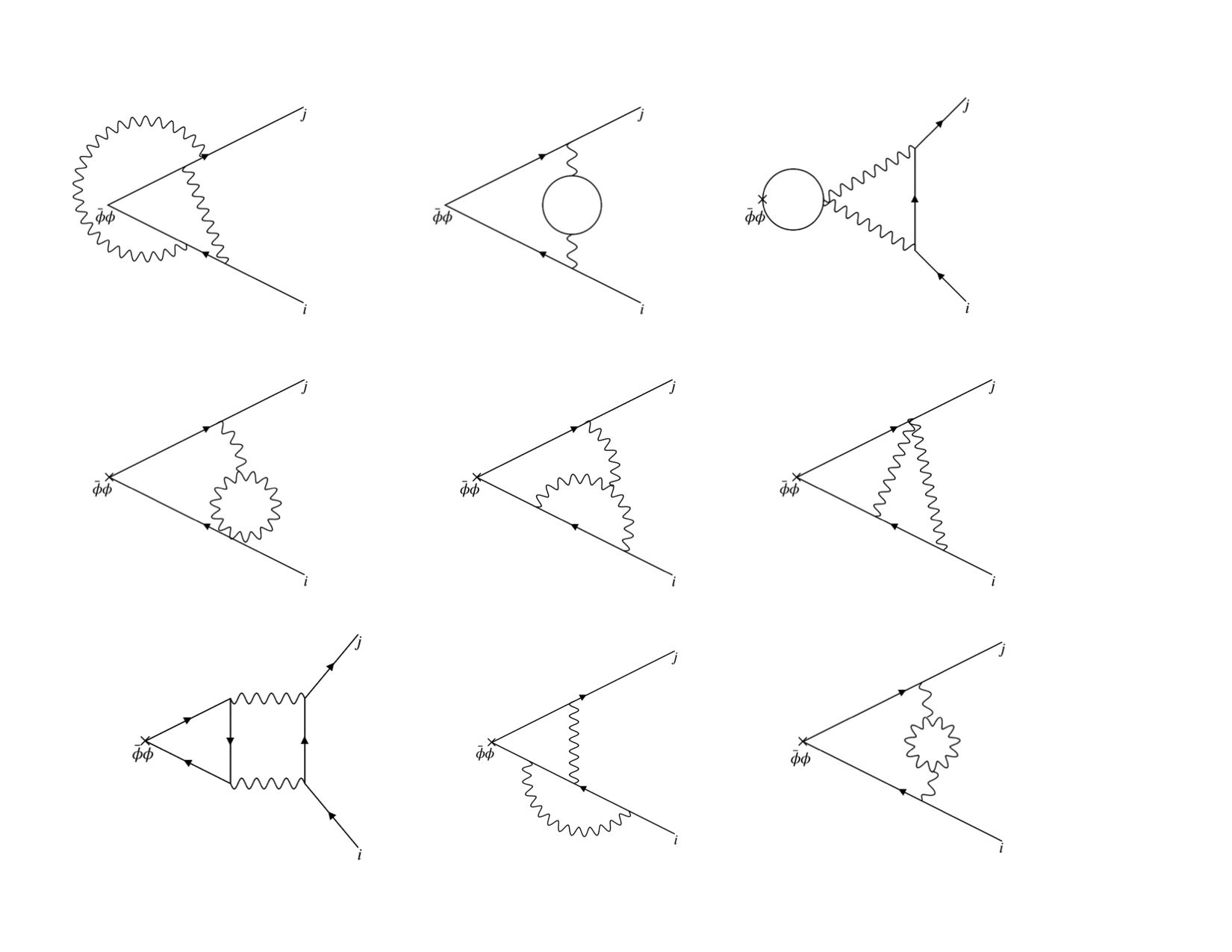}
  \caption{All the diagrams appearing in this figure do not contribute to the anomalous dimension at $1/N$.} \label{UVFin}
  \end{figure}

Apart from the diagrams depicted in fig.\ref{B1f} there are other
two-loop diagrams which do not contribute to the anomalous dimension
at order $1/N$.  
They are depicted in Figure \ref{UVFin}.  
  
\begin{figure}[h]
\centering
\includegraphics[scale=0.15]{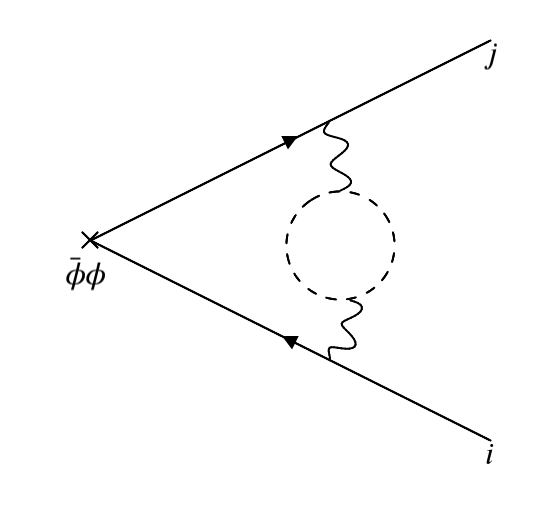}
\caption{Diagram with a ghost loop which cancels the last diagram in Fig.\ref{UVFin}.}\label{gh}
 \end{figure}
 \newpage
\subsection{Critical Fermionic Theory}
\begin{figure}
    \centering
   \includegraphics[scale=0.1]{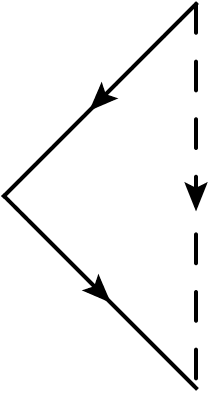}
   \hspace{0.3cm}\includegraphics[scale=0.1]{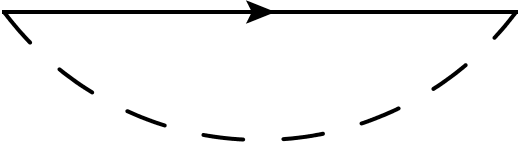}
         \includegraphics[scale=0.1]{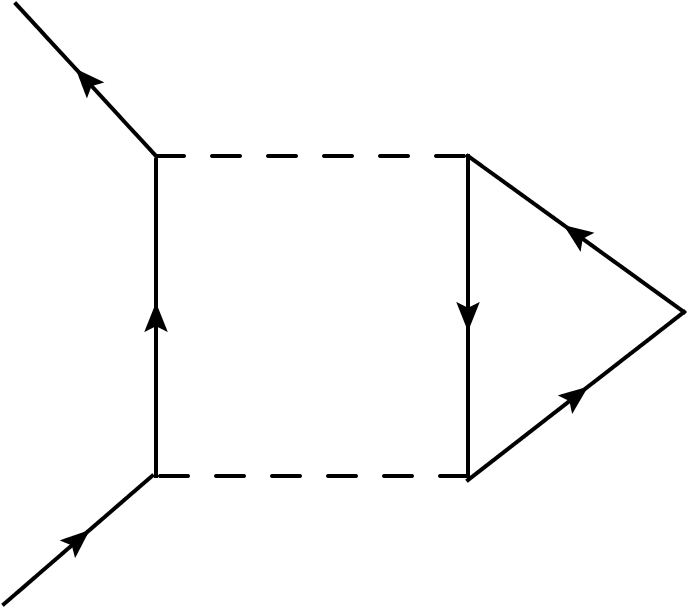}
         \includegraphics[scale=0.1]{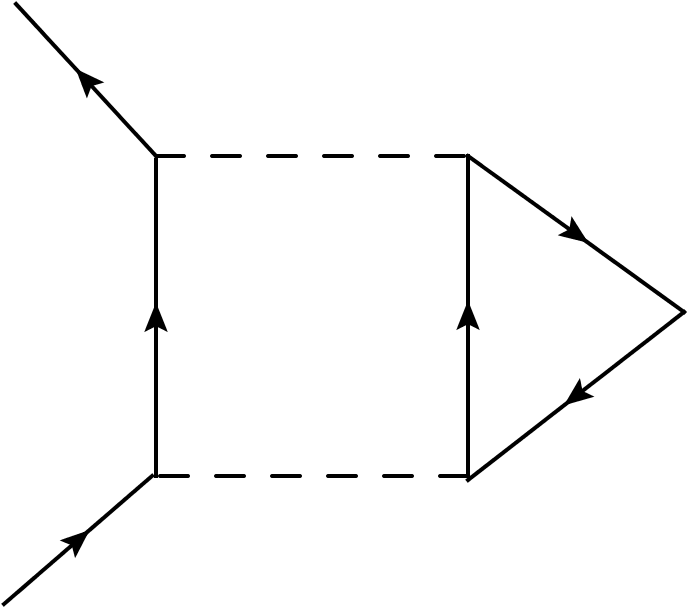}
    \caption{The feynman diagrams that contribute  in the critical fermionic theory \cite{Giombi:2016zwa}.}
    \label{fig:CF1}
\end{figure}
The order $\lambda_f^2$ correction anomalous dimension in the critical fermionic theory was determined through a direct Feynman diagram computation in \cite{Muta:1976js, Giombi:2016zwa} to be
\begin{equation}
\gamma_0 = \frac{1}{N_f}  \left( -\frac{16}{3 \pi ^2 }+ \frac{4}{9} \lambda_f^2 + O(\lambda_f^4) \right).\label{two-loop-critical-fermionic}
\end{equation}
The diagrams that contribute to the anomalous dimension at order $\lambda_f^2$ in the critical fermionic theory are depicted in Fig. \ref{fig:CF1}.
\section{Quasi-Fermionic Theory}\label{QFT}

The leading order $1/N$ anomalous dimension for the critical bosonic theory appears in \cite{Lang:1992zw} (see also \cite{Skvortsov:2015pea, Giombi:2016hkj}). We carried out a calculation of the order-$\lambda_b^2$ correction to this quantity to obtain:
\begin{equation}
\tilde{\gamma}_0 = \frac{1}{N_b}  \left( -\frac{16}{3 \pi ^2 }+ \frac{4}{9} \lambda_b^2 + O(\lambda_b^4) \right) \label{two-loop-critical-bosonic}.
\end{equation}
To order $\lambda_f^2$, the anomalous dimension of $\tilde{j}_0$ in the regular fermionic theory was computed through a direct Feynman diagram technique in \cite{Gurucharan:2014cva, Giombi:2016zwa}
\begin{equation}
\tilde{\gamma}_0 =\frac{1}{N_f} \left( -\frac{4}{3} \lambda_f^2 + O(\lambda_f^4)\right). \label{two-loop-fermionic}
\end{equation}

\begin{figure}[h]
   \includegraphics[scale=0.1]{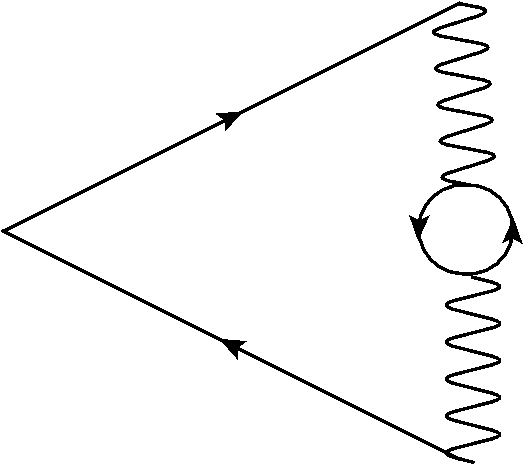}
   \hspace{0.3cm}\includegraphics[scale=0.1]{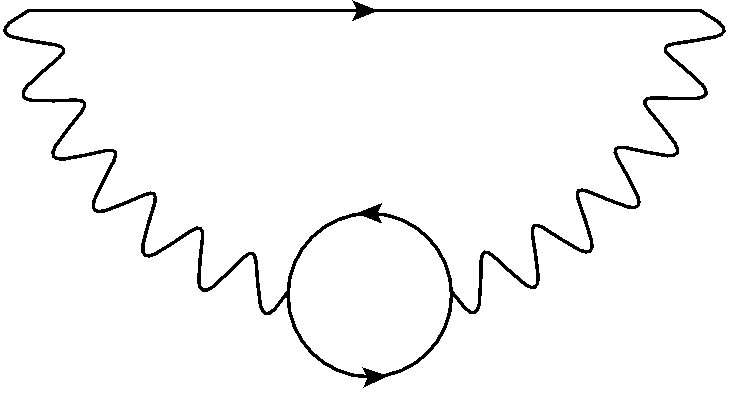}
         \includegraphics[scale=0.1]{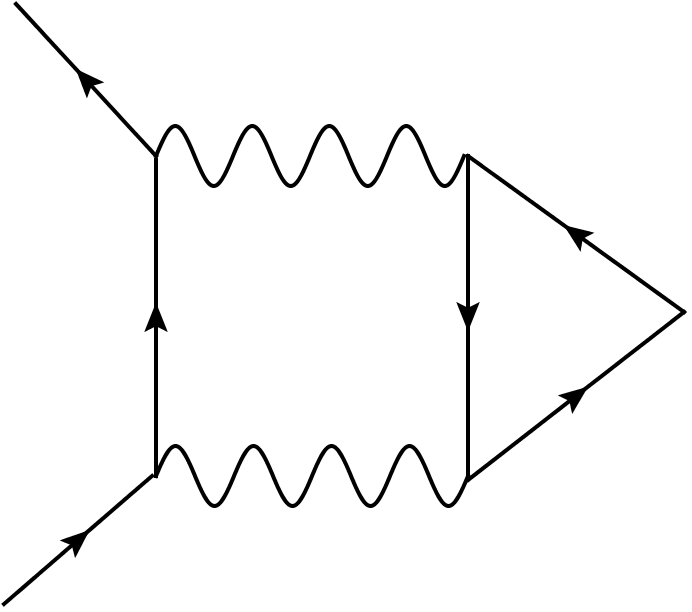}
         \hspace{0.2cm}\includegraphics[scale=0.1]{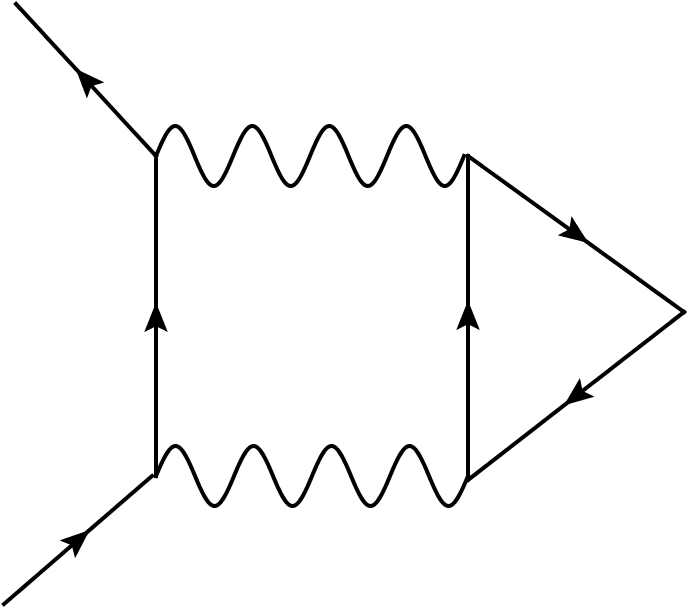}
    \centering
\caption{The feynman diagrams that contribute  in the Regular fermionic theory \cite{Giombi:2016zwa}.}
    \label{fig:RF1}
\end{figure}
The diagrams that contribute to the anomalous dimension at order $\lambda_f^2$ in the regular fermionic theory are depicted in Fig. \ref{fig:RF1}.
\section{All loop test}\label{ALT}
Here we briefly review the computation in \cite{Aharony:2018pjn} where the authors determine the $1/N$ correction to the sum of the anomalous dimension in the quasi-bosonic and the quasi-fermionic theories which will later serve as a highly non-trivial all loop consistency check for our conjecture.

To this end, the authors begin by the action of critical boson $(S_{C B}(\phi, \sigma))$ and the regular fermionic theories are  given by
\begin{widetext}
	\begin{align}
	S_{C B}\left(\phi, \sigma_{B}\right)&=\int d^{3} x[ i \varepsilon^{\mu \nu \rho} \frac{\kappa_{B}}{4 \pi} \operatorname{Tr}\left(A_{\mu} \partial_{\nu} A_{\rho}-\frac{2 i}{3} A_{\mu} A_{\nu} A_{\rho}\right)+i \varepsilon^{\mu \nu \rho} \frac{N_{B} \kappa_{B}^{\prime}}{4 \pi} B_{\mu} \partial_{\nu} B_{\rho} \notag\\ &~~~~~~~~~~~~~~~\left.+D_{\mu} \bar{\phi} D^{\mu} \phi+\sigma_{B} \bar{\phi} \phi\right] \\
	S_{R F}(\psi)&=\int d^{3} x\left[i \varepsilon^{\mu \nu \rho} \frac{\kappa_{F}}{4 \pi} \operatorname{Tr}\left(A_{\mu} \partial_{\nu} A_{\rho}-\frac{2 i}{3} A_{\mu} A_{\nu} A_{\rho}\right)+\bar{\psi} \gamma_{\mu} D^{\mu} \psi\right]
	\end{align}
\end{widetext}
where $k_B$ corresponds to the level of $SU(N_B)$  gauge field denoted above as $A_{\mu}$ and $k'_B$ corresponds to the level of a $U(N_B)$ gauge field  $B_{\mu}$. These two combine to form $U\left(N_{B}\right)=\left(S U\left(N_{B}\right) \times U(1)\right) / Z_{N_{B}}$.

The action for the regular bosonic and critical fermionic theories may be obtained as a relevant deformation of the critical bosonic and regular fermionic theories as follows
\begin{align}
 S_{R B}(\phi, \sigma, \zeta)&=S_{C B}(\phi, \sigma)-\int \tilde{J}_{0}(x) \zeta(x)\nonumber\\&~~~~+\frac{(2 \pi)^{2}}{\kappa_{B}^{2}}\left(x_{6}^{B}+1\right) \int \zeta^{3}(x)\label{RB1}\\
 S_{C F}(\psi, \zeta)&=S_{R F}(\psi)-\int J_{0}^{F}(x) \zeta(x)r\\&~~~~+\frac{(2 \pi)^{2}}{\tilde{\kappa}_{F}^{2}} x_{6}^{F} \int \zeta^{3}(x)\label{CF1}
\end{align}
where  the subscripts $RB,CB,RF$ and $CF$ denote the regular bosonic theory, critical bosonic theory, regular fermionic theory and critical fermionic theory respectively. Note that $\tilde{J_0}$ and $J_0^{F}$ are operators of scaling dimension $2$ which act as  sources for  new dynamical field $\zeta$. $x_6^B$  and $x_6^F$ are parameters.

The effective action for the critical fermionic and the regular bosonic  theories is obtained by integrating out the appropriate fields as
\begin{align}
 \int D \phi D \sigma e^{-S_{R B}(\phi, \sigma, \zeta)}&=e^{-S_{R B}^{e f f}(\zeta)} \\ \int D \psi e^{-S_{C F}(\psi, \zeta)}&=e^{-S_{C F}^{\epsilon f f}(\zeta)}
\end{align}
The authors observe that for $x_6^B=x_6^F$ the theories in eq.(\ref{RB1}) and eq.(\ref{CF1}) are identical and the conjectured duality leads to
\begin{align}
 S_{R B}^{e f f}\left(\zeta ; \kappa_{B}, \lambda_{B}\right)=S_{C F}^{e f f}\left(\zeta ;-\kappa_{B}, \lambda_{B}-\operatorname{sgn}\left(\lambda_{B}\right)\right)
\end{align}
This in turn implies that 1PI quantum effective action for both the theories are also identical as they are computed through the path integral of the above effective actions
\begin{align}
 S_{R B}^{1 P I}(\zeta)=S_{C F}^{1 P I}(\zeta)
\end{align}
In order to extract the difference between the anomalous dimensions of the two theories, the authors first evaluate the UV cut-off ($\Lambda$) dependent quintic and quartic terms in effective action $S_{R B}^{e f f}(\zeta)$ at leading order in $1/N$ to be
\begin{widetext}
\begin{align}
 S_{R B}^{e f f}&=\frac{g_{2}}{2 \kappa_{B}} \int \frac{d^{3} q}{(2 \pi)^{3}}|q|\left(\frac{|q|}{\Lambda}\right)^{\frac{2 \delta_{B}\left(\lambda_{B}\right)}{\kappa_{B}}} \zeta(q) \zeta(-q)
 \notag\\ &+\frac{g_{3}}{6 \kappa_{B}^{2}} \int \frac{d^{3} q_{1}}{(2 \pi)^{3}} \frac{d^{3} q_{2}}{(2 \pi)^{3}} \frac{d^{3} q_{3}}{(2 \pi)^{3}}(2 \pi)^{3} \delta\left(q_{1}+q_{3}\right) \zeta\left(q_{1}\right) \zeta\left(q_{2}\right) \zeta\left(q_{3}\right) \notag\\ &+\frac{\tilde{g}_{3}}{6 \kappa_{B}^{2}} \int \frac{d^{3} q_{1}}{(2 \pi)^{3}} \frac{d^{3} q_{2}}{(2 \pi)^{3}} \frac{d^{3} q_{3}}{(2 \pi)^{3}}(2 \pi)^{3} \frac{\delta_{B}}{\kappa_{B}} \ln \left(\frac{\Lambda}{\left|q_{1}\right|+\left|q_{2}\right|+\left|q_{3}\right|}\right) \delta\left(q_{1}+q_{2}+q_{3}\right) \zeta\left(q_{1}\right) \zeta\left(q_{2}\right) \zeta\left(q_{3}\right)\notag 
\\&-\frac{1}{24 \kappa_{B}^{3}} \int \prod_{i=1}^{4} \frac{d^{3} q_{i}}{(2 \pi)^{3}}(2 \pi)^{3} \delta\left(q_{1}+q_{2}+q_{4}+q_{4}\right) \kappa_{B}^{3} \tilde{G}_{4}^{0}\left(q_{1}, q_{2}, q_{3}, q_{4}\right) \zeta\left(q_{1}\right) \zeta\left(q_{2}\right) \zeta\left(q_{3}\right) \zeta\left(q_{4}\right)\notag\\& -\frac{1}{5 ! \kappa_{B}^{4}} \int \prod_{i=1}^{5} \frac{d^{3} q_{i}}{(2 \pi)^{3}}(2 \pi)^{3} \delta\left(q_{1}+q_{2}+q_{3}+q_{4}+q_{5}\right) \kappa_{B}^{4} \tilde{G}_{5}^{0}\left(q_{1}, q_{2}, q_{3}, q_{4}, q_{5}\right) \zeta\left(q_{1}\right) \zeta\left(q_{2}\right) \zeta\left(q_{3}\right) \zeta\left(q_{4}\right) \zeta\left(q_{5}\right)\label{SEf}
\end{align}
\end{widetext}
where $g_2,g_3$ and $\tilde{g_3}$ are functions of $\lambda_B$ and $ x_6^B$ (See \cite{Aharony:2018pjn} for details). Following this the authors then compute the one loop correction to the above 
\begin{align}
 S_{R B}^{1 P I}(\zeta)-S_{R B}^{e f f}(\zeta)=\frac{1}{2 \kappa_{D}^{2}} \int \frac{d^{3} q}{(2 \pi)^{3}} \delta \Gamma_{2}(q) \zeta(q) \zeta(-q)
\end{align}
where
\begin{align}
\delta \Gamma_{2}(q)=-\frac{32}{3 \pi \sin \pi \lambda_{B}}|q| \ln \left(\frac{\Lambda}{|q|}\right)
\end{align}
Hence  the 1PI quantum effective action for the regular bosonic theory in the above equation is of the form
\begin{align}
 S_{R B}^{1 P I}=\frac{g_{2}}{2 \kappa_{B}} \int \frac{d^{3} q}{(2 \pi)^{3}}|q|\left(\frac{|q|}{\Lambda}\right)^{\frac{2 \delta^{\prime}_\beta\left(\lambda_{B}\right)}{\kappa_{B}}} \zeta(q) \zeta(-q)\label{SPI}
\end{align}
Comparing  eq.(\ref{SEf}) and eq.(\ref{SPI}) we obtain the difference between the anomalous dimension of the two theories to be
\begin{align}
 \delta_{B}^{\prime}-\delta_{B}=\frac{16}{3 \pi \sin \pi \lambda_{B}}\label{delt}
\end{align}
These are related to the scaling dimension of the $\zeta$ and $\sigma$ operators  $\Delta_\zeta=\Delta_{j_0}$ and $\Delta_\sigma=\Delta_{\tilde{j}_0}$ as
\begin{align}
 \Delta_{\zeta}=1-\frac{\delta_{B}^{\prime}}{\kappa_{B}}~~~~~~\Delta_{\sigma}=2+\frac{\delta_{B}}{\kappa_{B}}
\end{align}
Comparing the above equation and eq.(\ref{delt}) we obtain
\begin{align}
 \delta_{B}^{\prime}=-\gamma_0\notag\\
\delta_{B}=\tilde{\gamma}_0.\label{delg}
\end{align}
Hence, eq.(\ref{delt}) and eq.(\ref{delg}) lead to the following relation between the anomalous dimension of the quasi-bosonic and quasi-fermionic theories 
\begin{align}
 \tilde{\gamma}_0+\gamma_0&=-\frac{16\lambda_b}{3\pi\sin \pi \lambda_b} \frac{1}{N_b}=-\frac{16\lambda_f}{3\pi\sin \pi \lambda_f} \frac{1}{N_f} \notag\\&=
-\frac{32}{3 \pi^2} \frac{1}{\tilde{N}}, \label{translate}
\end{align}
Note the above relation at  order $\lambda^2$ becomes
\begin{align}
\tilde{\gamma}_0+\gamma_0=-\frac{16}{3 \pi ^2}-\frac{8 \lambda_b ^2}{9}+O[\lambda_b^4]
\end{align}
Note  that the above result is exaclty satisfied by the two loop perturbative expressions for the anomalous dimension of the regular and critical bosonic theories given by eq.(\ref{anmcal}) and eq.(\ref{two-loop-critical-bosonic}). Similarly its easy to check the  above equation is satified by the result for the regular and  critical fermionic theories in  eq.(\ref{two-loop-fermionic}) and eq.(\ref{two-loop-critical-fermionic}). 
In the subsequent sections we will demonstrate that our conjectured expression for the anomalous dimension satisfies  \eqref{translate} to all orders in $\lambda$. This will provide strong evidence for the consistency of our conjecture.

\section{Towards an all loop result}\label{kr}

The authors in \cite{Giombi:2016zwa} utilized the non-conservation of the higher spin currents to demonstrate that $1/N$ higher-spin spectrum for the quasi-fermionic theory,to all orders in $\tilde{\lambda}_{QF}$ for the spinning operator $J_s$, is given by 
\begin{equation}\label{anom-s}
\gamma^{QF}_s =\frac{1}{\tilde{N}}\left( a_s^{QF}  \frac{\tilde{\lambda}_{QF}^2}{1+\tilde{\lambda}_{QF}^2} + b_s^{QF}  \frac{\tilde{\lambda}_{QF}^2}{(1+\tilde{\lambda}_{QF}^2)^2}\right)+O(\frac{1}{N^2}).
\end{equation}
Here $\gamma^{QF}_s = \Delta_s-(s+1)$ is the anomalous dimension of the spin-$s$ primary. A similar expression holds for the quasi-bosonic theory.

The expressions for the spin-dependent constants turn out to be identical\footnote{This is an unexplained coincidence at present, and is not true at order $1/N^2$, as can be seen from \cite{Manashov:2016uam, Manashov:2017xtt}.} for both the quasi-bosonic and quasi-fermionic theories and is

\begin{eqnarray}
  a_s & = & \begin{cases} \frac{16}{3\pi^2} \frac{s-2}{2s-1}\,, & \text{for even $s$}\,, \\
 \frac{32}{3\pi^2} \frac{s^2-1}{4s^2-1}\,, &  \text{ for odd $s$}\,,\end{cases} \\
  \label{bs-eq-intro}
 b_s & = & \begin{cases}
 \frac{2}{3 \pi
   ^2} \left(3 \displaystyle \sum_{n=1}^s \frac{1}{n-\frac{1}{2}}  +f(s)\right)\,, & \\\text{for even $s$}\,, \\
    \frac{2}{3 \pi ^2} \left(3
  \displaystyle \sum_{n=1}^s \frac{1}{n-\frac{1}{2}}+g(s)\right)\,, &\\ \text{ for odd $s$}\,.
\end{cases}
\\\text{where}~ f(s)&=&\frac{-38 s^4+24 s^3+34 s^2-24
   s-32}{4 s^4-5 s^2+1} \nonumber
   \\\text{and}~  g(s)&=&\frac{20-38 s^2}{4
   s^2-1}\nonumber
\end{eqnarray}
\subsection{Failure of the naive analytic continuation}

Note that unlike the higher spin operators the $j_0$ operator is not a conserved current at large-$N$ and hence,  the above result does not apply to the case of spin-$0$. However, it serves as an inspiration for our conjecture. Naively, one might be tempted to ``analytically continue'' the expressions for $a_s$ and $b_s$ in \cite{Giombi:2016zwa}, to $s=0$, using 
\begin{equation}
  \sum_{n=1}^s \frac{1}{n-1/2}= \gamma -\psi(s)+2\psi(2s) = H_{s-1/2}+2 \ln 2 \,
\end{equation}
resulting in 
\begin{equation}
    a^{AC}_0 \rightarrow \frac{32}{3 \pi ^2},~~ b^{AC}_0 \rightarrow -\frac{64}{3 \pi ^2}.
\end{equation}
Hence, this so-called ``analytic continuation'' gives the following answer for the value of $\gamma_0$ 
\begin{align}
 \gamma_0=\frac{32 \tilde{\lambda}_{QF}^2 \left(\tilde{\lambda}_{QF} ^2-1\right)}{3 \pi ^2 \left(\tilde{\lambda}_{QF}^2+1\right)^2}
\end{align}
Substituting for $\tilde{\lambda}_{QF}$ and $\tilde{N}$ from eq.(\ref{par1}) and  expanding the above to order $\lambda_{f} ^2$ we obtain the following expression
\begin{align}
 \gamma_0=-\frac{4 }{3}\lambda _f^2\frac{1}{ N_f} +O\left(\lambda_f ^4\right)
\end{align}
Interestingly, although the above term matches with the result obtained from the two-loop calculation in the regular fermionic (bosonic) theory given in eq.(\ref{two-loop-fermionic}) and eq.(\ref{anmcal}), it leads to an incorrect prediction for $\gamma_0$ and $\tilde{\gamma}_0$ in the critical bosonic (fermionic) theories given in eq.(\ref{two-loop-critical-fermionic}) and eq.(\ref{two-loop-critical-bosonic}) at $\tilde{\lambda}\rightarrow \infty$. This leads us to conclude that the naive analytic continuation fails to determine $1/N$ correction for the anomalous dimension of scalar operators in critical theories.

\section{Our conjecture}\label{Conj}
Here, we conjecture that the anomalous dimension of the scalar still takes the form given by equation \eqref{anom-s} for $s=0$
however, with different constants $a_0$ and $b_0$ than those obtained from the naive analytic continuation as follows\footnote{Notice that there is a characteristic double pole at $\tilde{\lambda}_{QB/QF}=\pm i$. It would be interesting to perform the analysis analogous to that in \cite{Gurucharan:2014cva} near this pole and examine whether this would lead towards a proof of our conjecture. We thank the anonymous referee for making this interesting observation.} 
\begin{align}\label{anom-s2}
\tilde{\gamma}_0  =\frac{1}{\tilde{N}}\left( a_0^{QF}  \frac{\tilde{\lambda}_{QF}^2}{1+\tilde{\lambda}_{QF}^2} + b_0^{QF}  \frac{\tilde{\lambda}_{QF}^2}{(1+\tilde{\lambda}_{QF}^2)^2}\right)+O(\frac{1}{N^2})\notag\\
\gamma_0  =\frac{1}{\tilde{N}}\left( a_0^{QB}  \frac{\tilde{\lambda}_{QB}^2}{1+\tilde{\lambda}_{QB}^2} + b_0^{QB}  \frac{\tilde{\lambda}_{QB}^2}{(1+\tilde{\lambda}_{QB}^2)^2}\right)+O(\frac{1}{N^2})\notag
\end{align}
Here, we attempt to determine these constants  by comparing our proposed expressions above perturbatively with the results listed in section \ref{kr}.
\subsection{Conjecture in Quasi-Fermionic theory: The $\tilde{\gamma}_0$ }
We can determine $a_0$ and $b_0$ in the quasi-fermionic theory, by first expanding around $\tilde{\lambda}_{QF}=\infty$
\begin{equation}
\tilde{\gamma}_0 =\frac{1}{\tilde{N}} \left( a^{QF}_0 + \frac{b^{QF}_0-a^{QF}_0}{\tilde{\lambda}_{QF}^2} + O( \frac{1}{\tilde{\lambda}_{QF}^4})\right).
\end{equation}
In terms of $\lambda_b$ and $N_b$ this is given by
\begin{align}
\tilde{\gamma}_0 =\frac{1}{N_b}\left(\frac{a^{QF}_0}{2}-\frac{(a^{QF}_0-3~ b^{QF}_0)~ \pi^2  \lambda_b^2}{24 }+O(\lambda_b^4)\right)
\end{align}
We can now compare this the two-loop result from the critical bosonic theory \eqref{two-loop-critical-bosonic}. This yields:
\begin{eqnarray}
a^{QF}_0 &= & -\frac{32}{3\pi^2}  \\
b^{QF}_0 & = & 0. 
\end{eqnarray}
We thus obtain the following expression for $\tilde{\lambda}_{QF}$
\begin{equation}
\tilde{\gamma}_0  =  -\frac{32}{3\pi^2} \frac{\tilde{\lambda}_{QF}^2}{1+\tilde{\lambda}_{QF}^2} \frac{1}{\tilde{N}} \label{con1}
\end{equation}
Making a perturbative expansion around $\tilde{\lambda}_{QF}=0$ which corresponds to $\lambda_{f}=0$ for the regular fermionic theory result, we find 
\begin{equation}
\tilde{\gamma}_0  = \frac{1}{N_f} \left( -\frac{4}{3} \lambda_f^2 + O(\lambda_f^4)\right).
\end{equation}
which precisely reproduces the two-loop result in the regular fermionic theory \eqref{two-loop-fermionic}, thus providing us a non-trivial test of our conjecture.
\subsection{Conjecture in Quasi-Bosonic theory: The $\gamma_0$ }
Repeating this procedure in the quasi-bosonic theory, we again find that 
\begin{eqnarray}
a^{QB}_0 &= & -\frac{32}{3\pi^2}  \\
b^{QB}_0 & = & 0. 
\end{eqnarray}
so
\begin{equation}
\gamma_0  =   -\frac{32}{3\pi^2} \frac{\tilde{\lambda}_{QB}^2}{1+\tilde{\lambda}_{QB}^2} \frac{1}{\tilde{N}}. \label{con2}
\end{equation}
Let us now expand the all loop expression above for  $\gamma_0$ from our conjecture around $\tilde{\lambda}_{QB}=0$ and $\tilde{\lambda}_{QB}=\infty$ to compare it with the results listed in section \ref{QBS1}. Expanding the expression in eq.(\ref{con2}) around $\tilde{\lambda}_{QB}=0$  re-expressed in terms of $\lambda_b$ yields
\begin{align}
\gamma_0  =\frac{1}{N_b} \left( -\frac{4}{3} \lambda_b^2+ O(\lambda_b^4)\right)
\end{align}
Notice that the above expression matches precisely with the anomalous dimension for regular bosonic theory we computed in section \ref{QBS1} given by eq.(\ref{anmcal}). Similarly expanding the expression in eq.(\ref{con2}) around $\tilde{\lambda}_{QB}=\infty$ re-expressed in terms of $\lambda_{f}$, we obtain
\begin{align}
\gamma_0 = \frac{1}{N_f}  \left( -\frac{16}{3 \pi ^2 }+ \frac{4}{9} \lambda_f^2 + O(\lambda_f^4) \right).
\end{align}
Once again this exactly matches with the result given by eq.(\ref{two-loop-critical-fermionic}) listed in section \ref{QBS1}. Hence, our conjecture exactly reproduces all known perturbative results for the anomalous dimension of the scalar primaries in both  quasi-bosonic and quasi fermionic theories which are available in the literature. Furthermore it also exactly reproduces the new computation we performed for the anomalous dimension in the regular bosonic theory.  Having established our conjecture in the perturbative regime, in the following subsection we provide a highly non-trivial all loop check for our conjecture.

\subsection{All loop check of our conjecture }
As a final non-trivial check of our conjecture we note that, using our expressions for $\gamma_0$ and $\tilde{\gamma}_0$, 
 we obtain
 \begin{align}
\tilde{\gamma}_0+\gamma_0 
&= -\frac{16\lambda_b}{3\pi\sin \pi \lambda_b} \frac{1}{N_b} = -\frac{16\lambda_f}{3\pi\sin \pi \lambda_f} \frac{1}{N_f}\notag\\& =
-\frac{32}{3 \pi^2} \frac{1}{\tilde{N}}, \label{translate1}
\end{align} which is exactly 
equation \eqref{translate} and is satisfied to all orders in $\lambda$. 
\subsection{ Anomalus dimension interms of $\lambda_b$ and $\lambda_f$ variables}
Let us conclude by presenting the expression for $\tilde{\gamma}_0$  and $\gamma_0 $ in terms of variables  $\lambda_b$ and $\lambda_f$ variables. Using \eqref{par1}, we have:
\begin{equation}\label{gt0}
\tilde{\gamma}_0=-\frac{8 \lambda_b}{3 \pi N_b }\cot\left(\frac{\pi \lambda_b}{2}\right)=-\frac{8 \lambda_f}{3 \pi N_f }\tan\left(\frac{\pi \lambda_f}{2}\right)
\end{equation}
and 
\begin{equation}\label{g0}
\gamma_0=-\frac{8 \lambda_b}{3 \pi N_b }\tan\left(\frac{\pi \lambda_b}{2}\right)=-\frac{8 \lambda_f}{3 \pi N_f }\cot\left(\frac{\pi \lambda_f}{2}\right).
\end{equation}
Note that our conjecture also reproduces the all the known results reported in sections I and II.

\section{Summary}
To summarize, we have proposed  a conjecture for the leading $1/N$ anomalous dimension of the scalar primary operator in $U(N)_k$ Chern-Simons theories coupled to a single fundamental field,
to all orders in $\lambda=\frac{N}{k}$. 
We demonstrated that our conjecture is consistent with all the existing two-loop perturbative results. We also performed a two-loop calculation of the anomalous dimension of the scalar primary $j_0$ in the bosonic theory, which provides an
additional test of our conjecture. Furthermore, we showed that our conjectured expression for the leading $1/N$ anomalous dimension for the quasi-bosonic and quasi-fermionic theories satisfies an all-loop relation that was previously derived in the literature.  This non-trivial consistency check gives  further evidence for our proposal.
\section*{Acknowledgements}
We would like to thank S Minwalla, O Aharony, A Gadde, T Hartman for fruitful discussions. SJ thanks TIFR, Mumbai for hospitality where part of the work was completed.
SP thanks IISER Pune and IFT, Madrid for hospitality where part of the work was completed. SP and SJ also thank the organizers of the Batsheva de Rothschild Seminar on Avant-garde methods for quantum field theory and gravity, for hospitality. SP acknowledges the support of a DST INSPIRE faculty award, and DST-SERB grant: MTR/2018/0010077. AM would like to acknowledge the support of CSIR-UGC (JRF) fellowship (09/936(0212)/2019-EMR-I). 
Research of SJ and VM is supported by Ramanujan Fellowship.  Finally, we would like to acknowledge our debt to the steady support of the people of India for research in the basic sciences. 

\begin{appendix}\label{fey}

\section{Conventions and Feynman Rules}
The Lagrangian is given by 
\begin{eqnarray}
S&=&S_{CS}+S_{RB} \\
S_{RB}&=&\int d^3x ~|D_{\mu} \phi|^2+\frac{\lambda_6}{3! N^2}(\phi^{\dagger}\phi)^3\\
S_{CS}&=&\frac{i k}{4 \pi}\int d^3x ~~\Tr(A\wedge dA+\frac{2}{3}A\wedge A\wedge A)
\\ &= & \frac{i k}{8 \pi}\int d^3x~ \epsilon_{\mu\nu\lambda}\big[A_{\mu}^a\partial_\nu A_{\lambda}^a-\frac{i}{3}f^{abc} A_{\mu}^aA_{\nu}^bA_{\lambda}^c\big]\nonumber
\end{eqnarray}
In expanding the Chern-Simons action, we used $\Tr(T^a T^b)=\frac{1}{2}\delta_{ab}$ as our convention for group generators. We will express all the divergent diagrams that contribute to the anomalous dimension in terms of $C_1$, $C_2$ and $C_3$ which are defined by following relations
\begin{eqnarray}
\Tr(T^aT^b) &= &\delta_{ab}C_1 \\
f^{acd}f^{bcd}&=&\delta^{ab} C_2\\
T^aT^a&=&I C_3. 
\end{eqnarray} 
In the normalization that we have chosen for $SU(N)$ generators,
\begin{equation}
C_1=\frac{1}{2}, ~~~C_2=-N, ~~~C_3=\frac{1}{2}(N-\frac{1}{N}).
\end{equation}

\begin{figure}
\begin{center}
	    \includegraphics[scale=0.5]{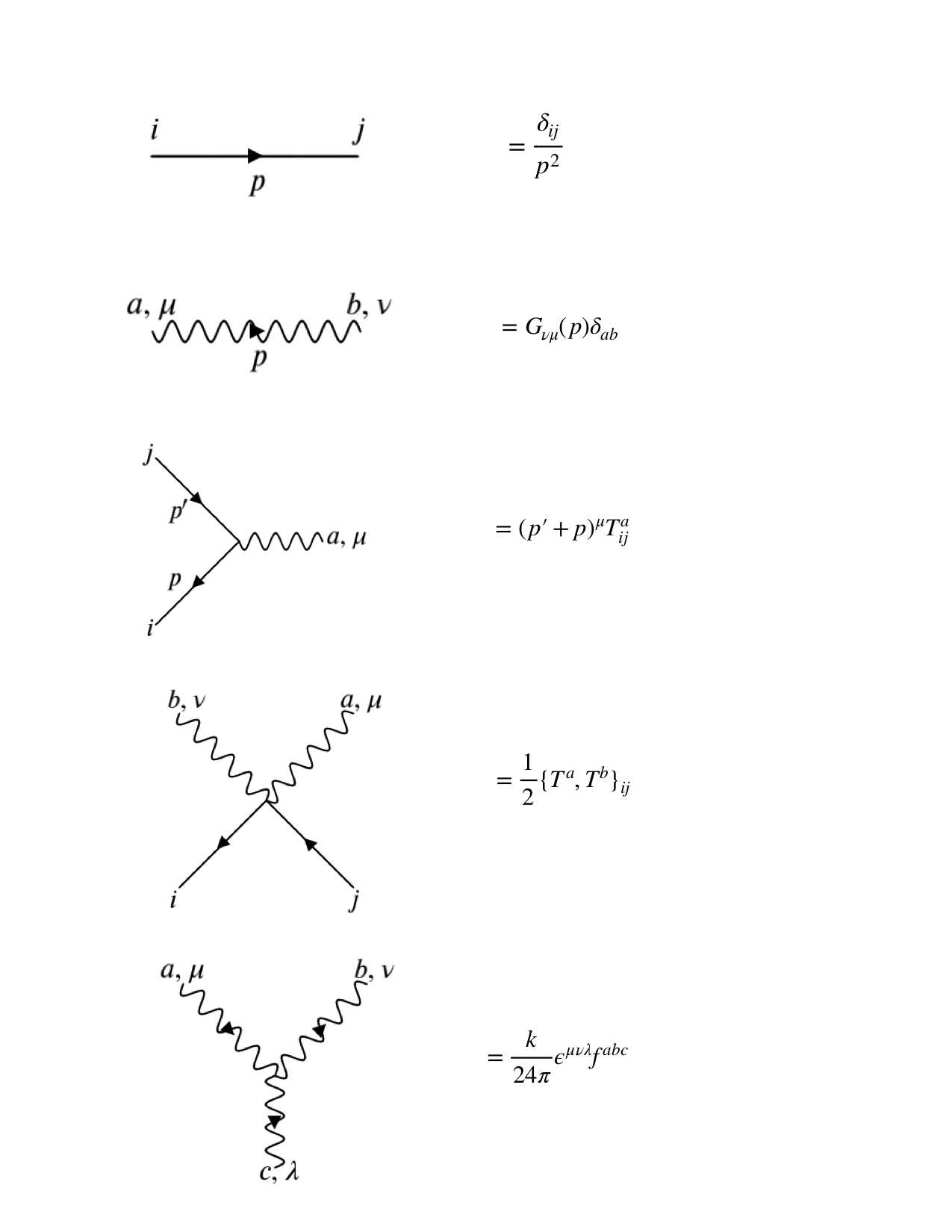}
\end{center}
\caption{Feynman rules for regular bosonic theory}
\label{Fig6}
\end{figure} 
We obtain the Feynman rules as depicted in Fig. \ref{Fig6}. If we work in the Landau gauge, as in \cite{Aharony:2011jz}, then we have gluon propagator to be as
\begin{equation}
G_{\mu\nu}=-\frac{4\pi}{k}\frac{\epsilon^{\mu\nu\delta}p_{\delta}}{p^2}.
\end{equation}

\section{Two-sided Pad\'e approximation}
Let us also observe that our conjecture can be thought of as a two-sided Pad\'e approximation. In this sense, even if our conjecture turns out to be incorrect, it provides a good estimate for the anomalous dimension of the scalar primary that takes into account all known weak-coupling and strong-coupling calculations.

Consider making an $(m,n)$-Pad\'e approximation of $\gamma_0$ as follows:
\begin{equation}
\gamma_0^{(m,n)} = \frac{A_0+A_2 \tilde{\lambda}_{QB}^2 + \ldots + A_{m} \tilde{\lambda}_{QB}^m}{1+B_2 \tilde{\lambda}_{QB}^2+\ldots+B_n \tilde{\lambda}^n}.
\end{equation}
We only include even powers of $\tilde{\lambda}$ as the anomalous dimension must be parity-invariant.

The $(2,2)$ Pad\'e approximation has three unknowns. We have four perturbative data to constrain it:
\begin{itemize}
  \item The fact that $\gamma_0$ vanishes when $\tilde{\lambda}_{QB}=0$.
  \item A two-loop calculation $\gamma_0$ in the regular-bosonic theory.
  \item The value of $\gamma_0$ in the critical fermionic theory at $\lambda_b=0$.
  \item A two-loop (order $\lambda_b^2$) calculation of $\gamma_0$ in the critical fermionic theory.
  \end{itemize}
Hence the Pad\'e-approximation is overconstrained. Nevertheless, it is possible to fit all four results with following choice of three coefficients.
\begin{equation}
A_0 = 0,~~ A_2 = -\frac{32}{3 \pi^2} , ~~B_2 = 1.
\end{equation}

Repeating the calculation to obtain a $(2,2)$ Pad\'e approximation for the quasi-fermionic theory, we obtain the same coefficients. However, we also have to impose the extra constraint of equation \eqref{translate}, which turns out to be  automatically satisfied.

Hence, the simplest Pad\'e approximation to the perturbative data we have seems to work very well. Of course, it is possible to obtain higher-order Pad\'e approximations that satisfy all these constraints, so our answer is not uniquely determined by this procedure. But, it is an interesting observation that, for a variety of physical quantities, such as planar three-point functions \cite{MZ}, planar four-point function of the scalar primary \cite{Turiaci:2018dht}, and the $1/N$ higher-spin spectrum \cite{Giombi:2016zwa}, a relatively simple Pad\'e approximation defined using the variables $\tilde{\lambda}$ and $\tilde{N}$, happens to coincide with the exact answer. 
 \end{appendix}

\bibliographystyle{JHEP}
\bibliography{CSBib}
\end{document}